\documentclass[review,number,sort&compress]{elsarticle}
 
\usepackage{lineno}
\usepackage{color}
\usepackage{ulem}
\usepackage{amsmath}
\usepackage{subfig}

\begin{document}

%\linenumbers

\begin{frontmatter}

\title{Correction Method for the Readout Saturation of the DAMPE Calorimeter}

\author[pmo]{Chuan Yue\corref{cor}}
\ead{yuechuan@pmo.ac.cn}

\author[pmo,ustcpmo]{Peng-Xiong Ma\corref{cor}}
\ead{mapx@pmo.ac.cn}

\author[leece1,leece2]{Margherita Di Santo}

\author[ustc]{Li-Bo Wu}

\author[gssi1,gssi2]{Francesca Alemanno}

\author[leece1,leece2]{Paolo Bernardini}

\author[gssi1,gssi2]{Dimitrios Kyratzis}

\author[pmo,ustcpmo]{Guan-Wen Yuan}

\author[pmo,ustcpmo]{Qiang Yuan}

\author[ustc]{Yun-Long Zhang}

\cortext[cor]{Corresponding author}
\address[pmo]{Key Laboratory of Dark Matter and Space Astronomy, Purple Mountain Observatory, Chinese Academy of Sciences, Nanjing 210008, China}
\address[ustcpmo]{School of Astronomy and Space Science, University of Science and Technology of China, Hefei 230026, China}
\address[leece1]{Dipartimento di Matematica e Fisica E. De Giorgi, Università del Salento, I-73100 Lecce, Italy}
\address[leece2]{Istituto Nazionale di Fisica Nucleare (INFN)–Sezione di Lecce, I-73100 Lecce, Italy}
\address[ustc]{State Key Laboratory of Particle Detection and Electronics, University of Science and Technology of China, Hefei 230026, China}
\address[gssi1]{Gran Sasso Science Institute (GSSI), Via Iacobucci 2, I-67100 L'Aquila, Italy}
\address[gssi2]{Istituto Nazionale di Fisica Nucleare (INFN) -Laboratori Nazionali del Gran Sasso, I-67100 Assergi, L'Aquila, Italy}

\begin{abstract}
The DArk Matter Particle Explorer (DAMPE) is a space-borne high energy cosmic-ray
and $\gamma$-ray detector which operates smoothly since the launch on 
December 17, 2015. The bismuth germanium oxide (BGO) calorimeter is one of 
the key sub-detectors of DAMPE used for energy measurement and electron-proton
identification. For events with total energy deposit higher than decades of 
TeV, the readouts of PMTs coupled on the BGO crystals would become 
saturated, which results in an underestimation of the energy measurement. 
Based on detailed simulations, we develop a correction method for the 
saturation effect according to the shower development topologies and 
energies measured by neighbouring BGO crystals. 
The verification with simulated and on-orbit events shows that this method can well 
reconstruct the energy deposit in the saturated BGO crystal.
\end{abstract}


\end{frontmatter}

\section{Introduction}
Measurements of the energy spectra of various cosmic ray (CR) nuclei are
the key to understanding the origin, propagation, and interaction of
these energetic particles \cite{Strong:2007nh, Amato:2017dbs, Gabici:2019jvz}. 
Current measurements carried out by magnetic spectrometer experiments reach 
very high precision up to TV rigidities \cite{Aguilar:2017hno}. At even higher 
energies, direct measurements by calorimeter experiments 
show interesting hints that the spectra of CR nuclei may have 
complicated structures \cite{Yoon:2017qjx, Atkin:2018wsp, Adriani:2019aft}.
However, these results are still subject to relatively large uncertainties, 
due to either limited statistics or large systematic uncertainties. 
Improved measurements are essential and necessary for addressing those
important questions of CR physics.

The DArk Matter Particle Explorer (DAMPE; \cite{ChangJ2014,DmpMission}) 
is an orbital mission for precision measurements of CR nuclei, electron/positrons, and 
$\gamma$-ray, supported by the strategic priority science and 
technology projects in space science of the Chinese Academy of Science. 
It was launched into a sun-synchronous orbit at an altitude of 500 km 
on December 17, 2015, and has been working smoothly for more than 4 
years since then. 
The scientific payload of DAMPE consists of four sub-detectors, 
including a Plastic Scintillator strip Detector (PSD; \cite{YuYH2017, Ding:2018lfn}), 
a Silicon-Tungsten tracKer-converter (STK; \cite{Azzarello2016, Tykhonov:2018stq}), 
a BGO imaging calorimeter (BGO; \cite{ZhangYL2012, ZhangZY2016}), and a NeUtron 
Detector (NUD; \cite{HuangYY2020}). These four sub-detectors work 
cooperatively to enable good measurements of charge, track, energy 
and particle-id of each incident particle \cite{Tykhonov:2017uno, Ma:2018brb, Dong:2019hnn}. 
Precise spectral measurements regarding electrons plus positrons \cite{DmpElectron} 
and protons \cite{DmpProton} in extended energy intervals, 
reveal interesting features and shed new light on the understandings of CR physics,
while improving the constraints on dark matter models 
\cite{Yuan:2017ysv, Pan:2018lhc, Yuan:2018rys, Yue:2019sxt}. 
The $\gamma$-ray identification technique \cite{XuZL2017} and analysis 
tool \cite{Duan:2019wns} have also been developed, with preliminary 
results \cite{2019ICRC...36..576L}.

The BGO calorimeter is the main sub-detector for energy measurement, 
which is designed as a total-absorption electromagnetic calorimeter 
of about 31.5 radiation length and 1.6 nuclear interaction length. It is 
composed of 14 layers, each layer consists of 22 BGO crystals 
($25 \times 25 \times 600$ mm$^{3}$) placed orthogonally in two 
dimensions \cite{ZhangYL2012}. The fluorescence signal of each BGO 
crystal is read out by two PMTs mounted on both ends. This design provides 
two independent energy measurements. Apart from measuring 
the energy deposits of the cascade showers produced by incident 
particles, the calorimeter images their shower developments, thereby 
serving as a hadron/lepton discriminator \cite{DmpElectron}.

At the very-high-energy end of DAMPE's capability, saturations of 
the low-gain readouts appear\footnote{{For protons 
and helium nuclei, the saturation may happen for deposited energies higher 
than $\sim20$~TeV.}}, which affect the precise measurement of the particle energy. 
For most of the saturated events, there are no more than one BGO crystal 
in the same layer showing the saturation effect. In this work, we develop 
a method to correct the saturated readout for those events, which is 
helpful in reconstructing the proper energy deposits of them. Applying such 
corrections would enable us to significantly enlarge the measurable 
energy ranges of CR nuclei.

\section{BGO Readout Saturation}
To fulfil the requirement of a wide energy coverage, from 5 GeV to 10 TeV 
for $e^{\pm}/{\gamma}$ and up to 100 TeV for nuclei, the scintillation 
light signal of each BGO crystal is read out from three different sensitive 
dynodes 2, 5, and 8 (Dy2, Dy5, and Dy8) of the PMTs, which corresponds to
low-gain, medium-gain, and high-gain channels, respectively \cite{ZhangZY2015}.
The response ratios of adjacent dynodes, i.e. Dy8/Dy5 and Dy5/Dy2, are 
carefully calibrated using high-energy shower events collected on orbit, 
which show good linear correlations and maintain stability over time
\cite{DmpCalibration}. Non-linearity effect from the conversion of the 
ionization energy to the light yield \cite{Kounine2017} has not been found 
for electrons up to a few TeV energies. However, for each PMT dynode, 
an upper limit of the ADC readout has been set beyond which the readout 
is discarded on orbit. The PMTs on the two ends of one BGO crystal 
(named S0 and S1) are coupled to the BGO bar with two different optical 
filters. The filter on the S1 end has a factor of $\sim$ 5 times 
attenuation with respect to the one on the S0 end, thereby the upper 
limit of the S1 end is about 5 times higher than that of the S0 end 
\cite{DmpMission}. Therefore, the energy deposit in each crystal can be measured 
two times independently for most of events, which is helpful to improve the energy 
resolution by combining the readouts from two ends. When the energy deposit 
in a BGO crystal is larger than the maximum measurable limit of S0 end, 
the energy deposit can still be properly reconstructed by the readout of S1 end after 
the attenuation correction \cite{WuLB2018}. However, when the energy deposit in 
one crystal is even larger than the maximum measurable limit of the S1 end, 
this event is defined as saturated and the energy information of this particular crystal is lost.

\begin{figure}[!ht]
  \centering  
  \includegraphics[width=.6\textwidth]{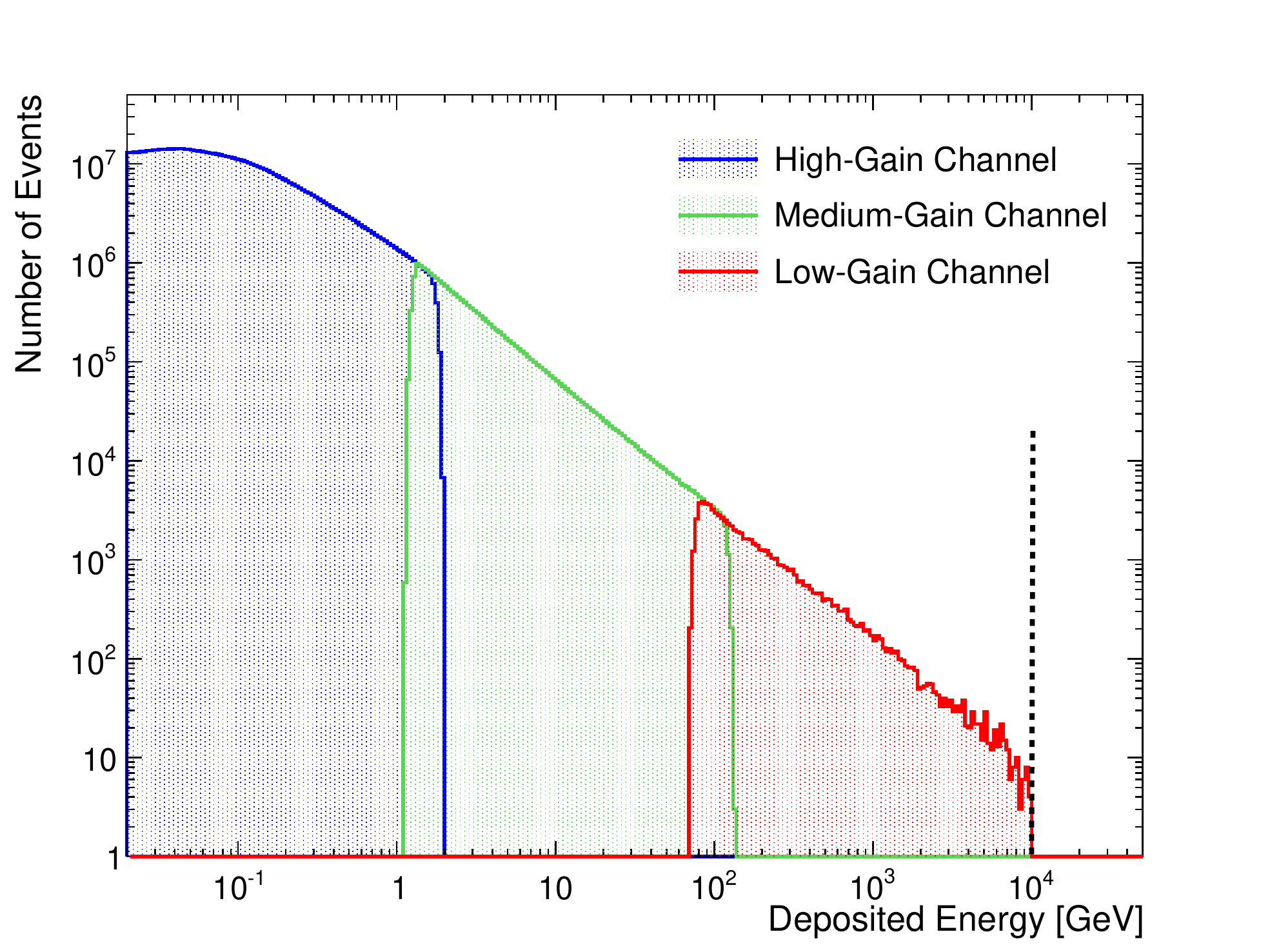}
  \caption{A typical energy deposit spectrum reconstructed from the S1 end of one BGO crystal. The blue, green and red histograms correspond to the high-gain (Dy8), medium-gain (Dy5), and low-gain (Dy2) ranges, respectively. The vertical black line represents the upper limit of the measurement.}
  \label{fig-1}
\end{figure}

\begin{figure}[!ht]
  \centering  
  \includegraphics[width=.8\textwidth]{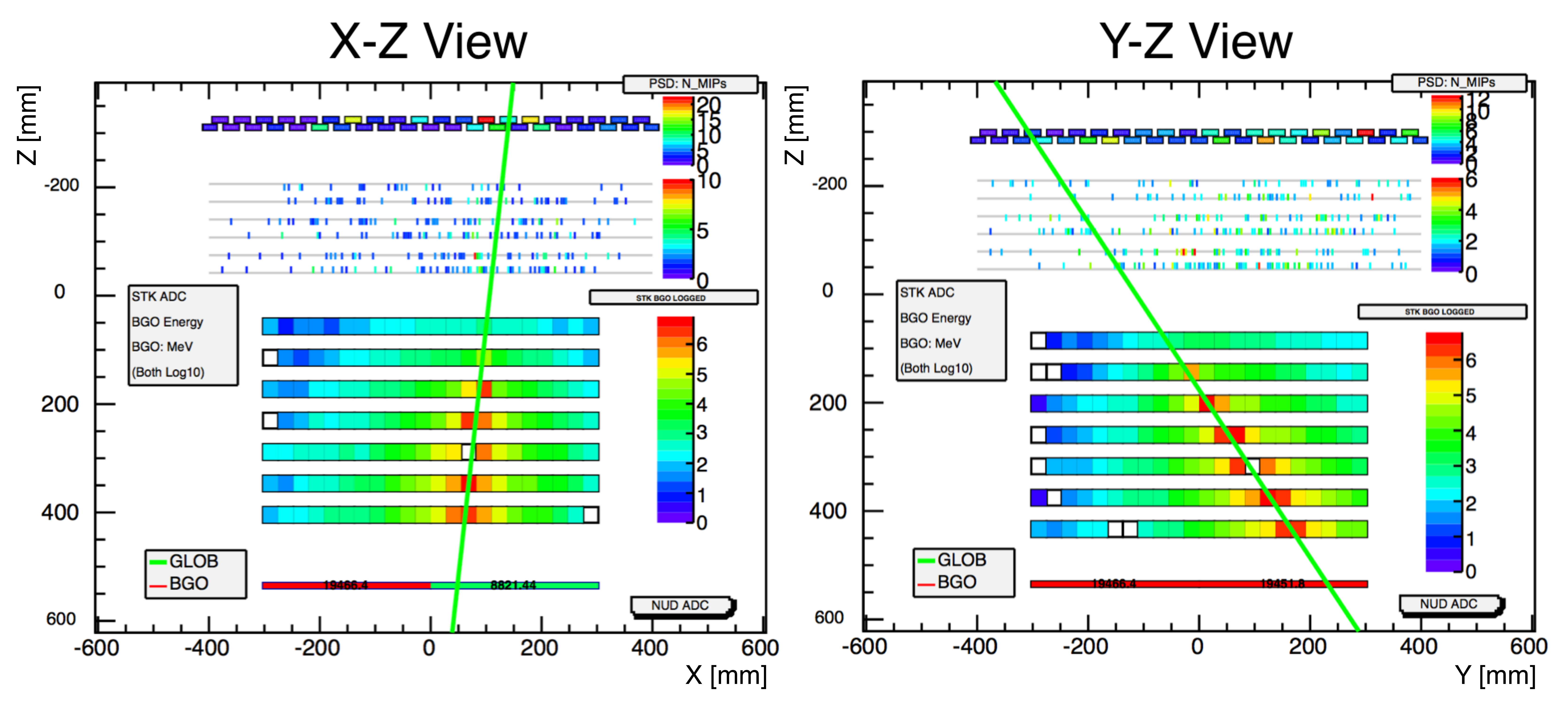}
  \caption{An illustration of a helium event with BGO readout saturation. The pre-correction total energy deposit is 49.4 TeV. The two empty BGO crystals on the shower axis are saturated, while the other empty crystals on the edge of shower are the ones without any deposited energy (or, the energy deposit is smaller than the noise threshold). }
  \label{fig-2}
\end{figure}

Fig.\ref{fig-1} shows a typical energy deposit spectrum reconstructed from the S1 end of one BGO crystal after the attenuation correction. A smooth transition between adjacent gain ranges can be clearly seen. The vertical black dashed line represents the upper measurement limit of the Dy2 readout channel, which is $\sim$10 TeV. As different PMTs have different gains \cite{ZhangZY2016}, the upper measurement limit of the S1 end varies from $\sim$4 TeV to $\sim$15 TeV. This upper limit is high enough for the measurement of $e^{\pm}/{\gamma}$ to energies of $\sim$10 TeV. However, for CR nuclei which are expected to be measures above energies of 100 TeV, the deposited energy in the calorimeter would exceed several tens of TeV, with the maximum energy in one single BGO bar exceeding several TeV. Therefore the saturation may appear for those very-high-energy events. Fig.\ref{fig-2} shows a helium event with saturation. The deposited energy is 49.4 TeV before correction. The actual deposited energy of this event should be much larger. 

\section{Method for the saturation correction}

\begin{figure}[!ht]
  \centering
  \includegraphics[width=.48\textwidth]{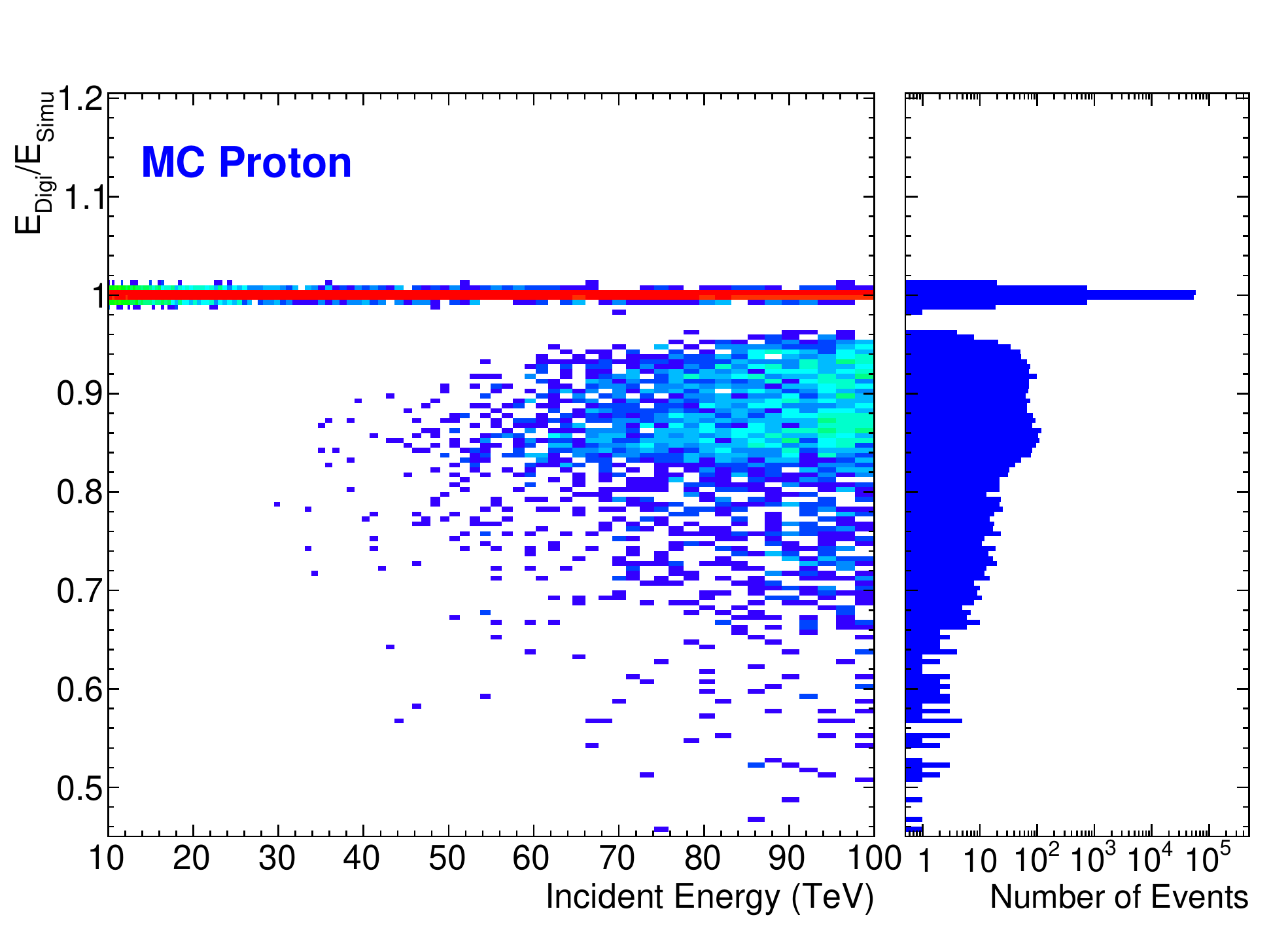}
  \includegraphics[width=.48\textwidth]{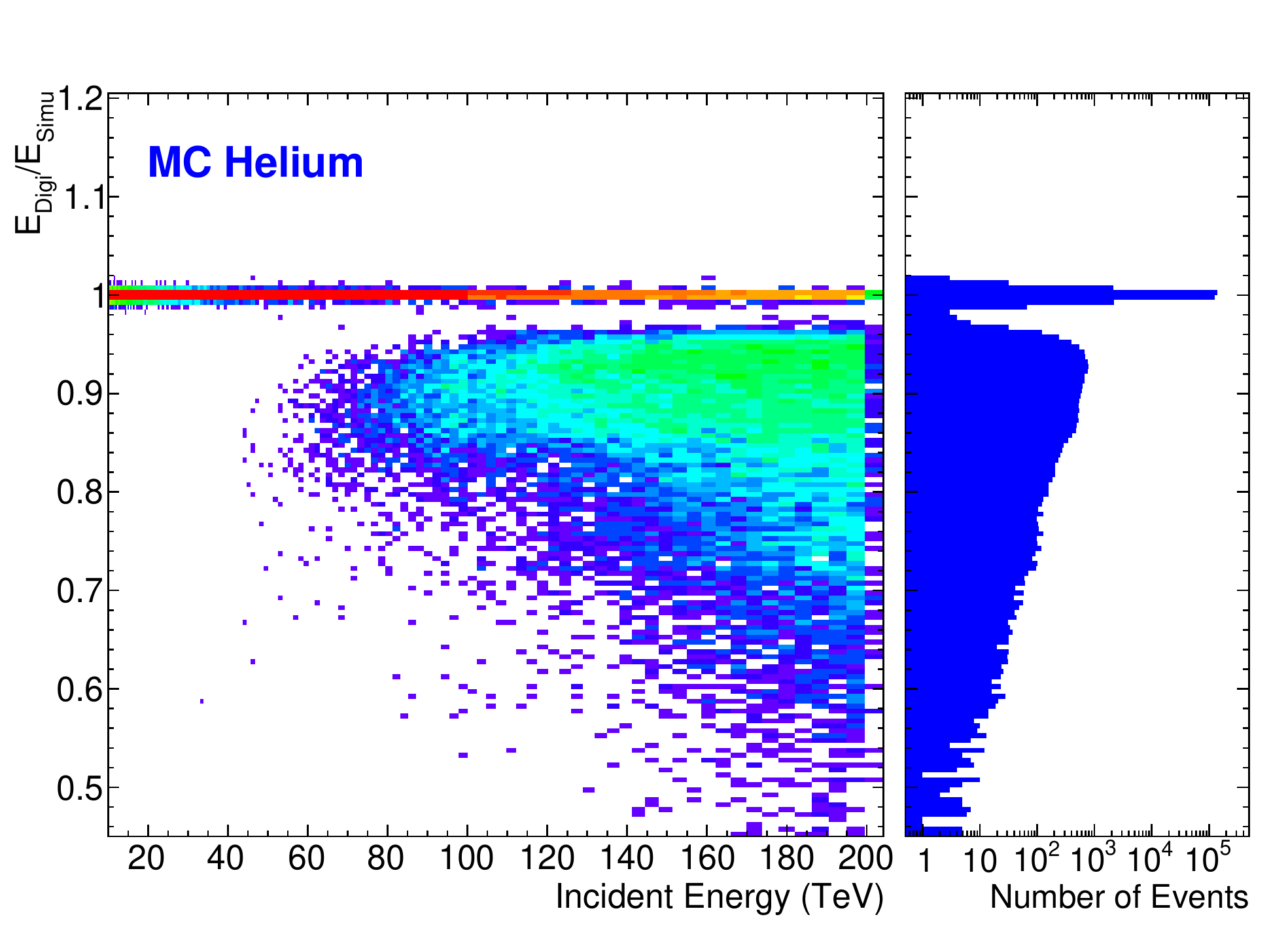}
  \caption{The ratios of digitized energy deposit to simulated energy deposit versus the incident energy for MC protons ({\tt left}) and helium nuclei ({\tt right}).}
  \label{fig-3}
\end{figure}

The saturation effect of the BGO readout has been taken into account in 
Monte Carlo (MC) simulation tool of DAMPE via importing saturation thresholds
in the digitization procedure \cite{YueC2017}. In this analysis, we use the
protons and helium nuclei sample generated with the FTFP\_BERT hadronic 
interaction physics list in the Geant4 software \cite{Allison2006}. 
Fig.\ref{fig-3} shows the ratios of digitized energy deposits 
($E_{\rm digi}$) to simulated energy deposits ($E_{\rm simu}$) for MC 
protons (left) and helium nuclei (right) with incident energies $\geq$10 TeV. 
The scattered points below 1 represent events that suffered from the readout 
saturation effect. The fraction of saturated events becomes higher with 
the increase of particle energies. Particularly, at 100 TeV of incident energy, the fraction 
of saturated events is $\sim$1.5\% ($\sim$1.2\%) for MC proton (helium).
Therefore, the saturation effect would be more and more important for spectral 
measurements of CRs at increasingly high energies.

For the flight data, there would be one or more saturated BGO crystal(s) for a single event, leading to a large discrepancy for energy measurement. Since we have lost the energy information of the saturated crystal, we need to estimate its energy deposit based on the other un-saturated crystals and the shower development information. By combining the energy information of neighbouring BGO crystals, we propose a two-step correction method to reconstruct the energy deposit(s) of the saturated crystal(s).

\begin{figure}[!ht]
  \centering  
  \includegraphics[width=.8\textwidth]{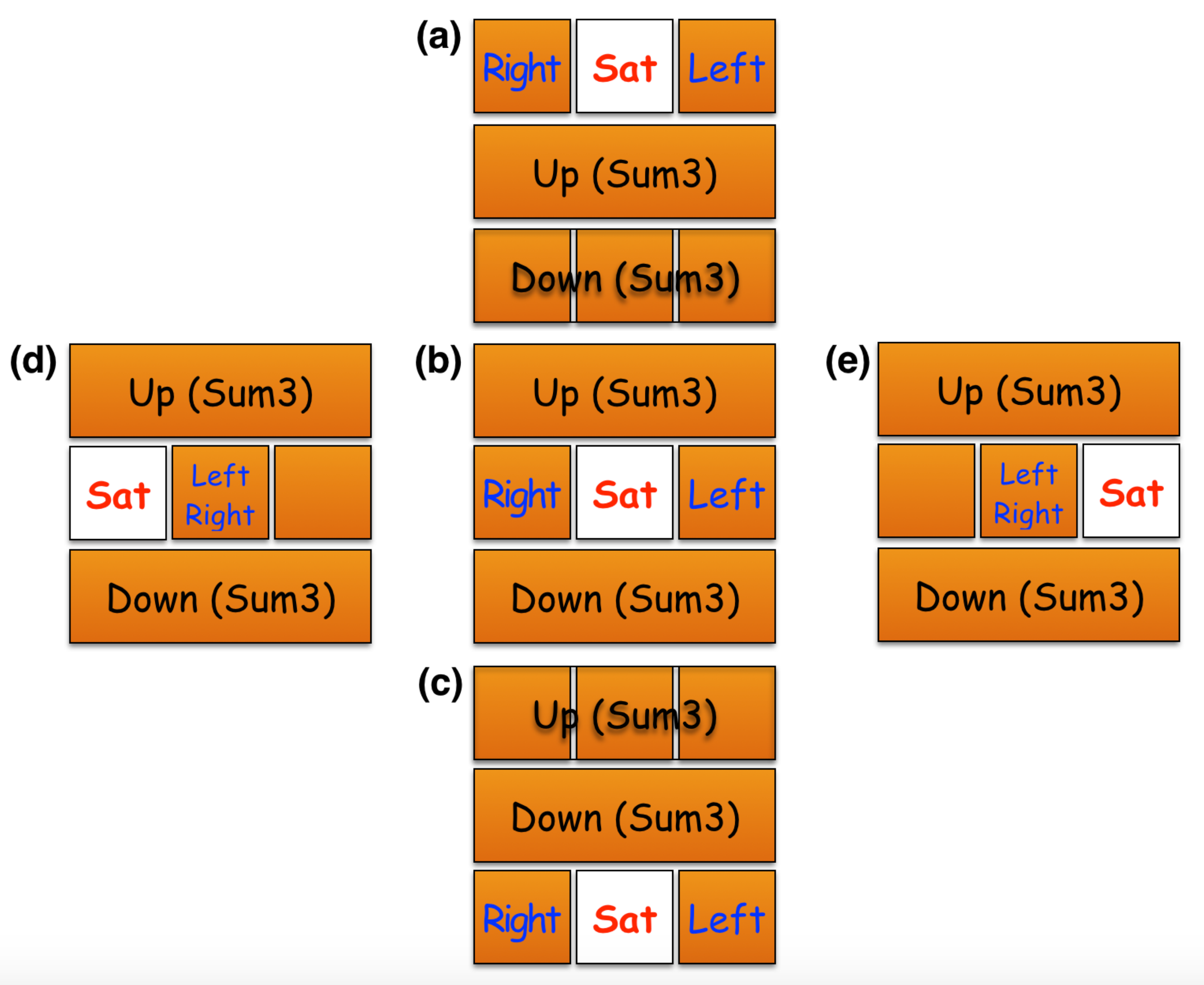}
  \caption{Classifications of events that need corrections: (a) for the top layer ($j=1$); (b) for middle layers ($j=2,...,13$); (c) for the bottom layer ($j=14$); (d) for saturated bar on the left edge; (e) for saturated bar on the right edge.}
  \label{fig-4}
\end{figure} 
 
\subsection{Left-right correction}
The simulations indicate that the saturated crystal should be the one with the maximum deposited energy in a certain BGO layer. As a prime estimation, we construct a correction variable $\eta_{\rm LR}$ based on the energies in the {\tt left} and {\tt right} neighbouring bars (see Fig.\ref{fig-4}), defined as follows:
%------Equation------
\begin{equation}
\eta_{{\rm LR},j} = \frac{E_{{\rm Max},j}}{E_{{\rm Max},j}+E_{{\rm Left},j}+E_{{\rm Right},j}},
\label{eq-etaLR}
\end{equation}
where $E_{{\rm Max},j}$ is the maximum energy deposit in the $j$th layer, $E_{{\rm Left},j}$ ($E_{{\rm Right},j}$) is the energy deposit in its left (right) neighbouring crystal. When the saturated bar is located on the edge of one layer (classes (d) and (e) in Fig.\ref{fig-4}), $E_{{\rm Right},j}$ or $E_{{\rm Left},j}$ are counted twice.

\begin{figure}[!ht]
  \centering
  \includegraphics[width=.48\textwidth]{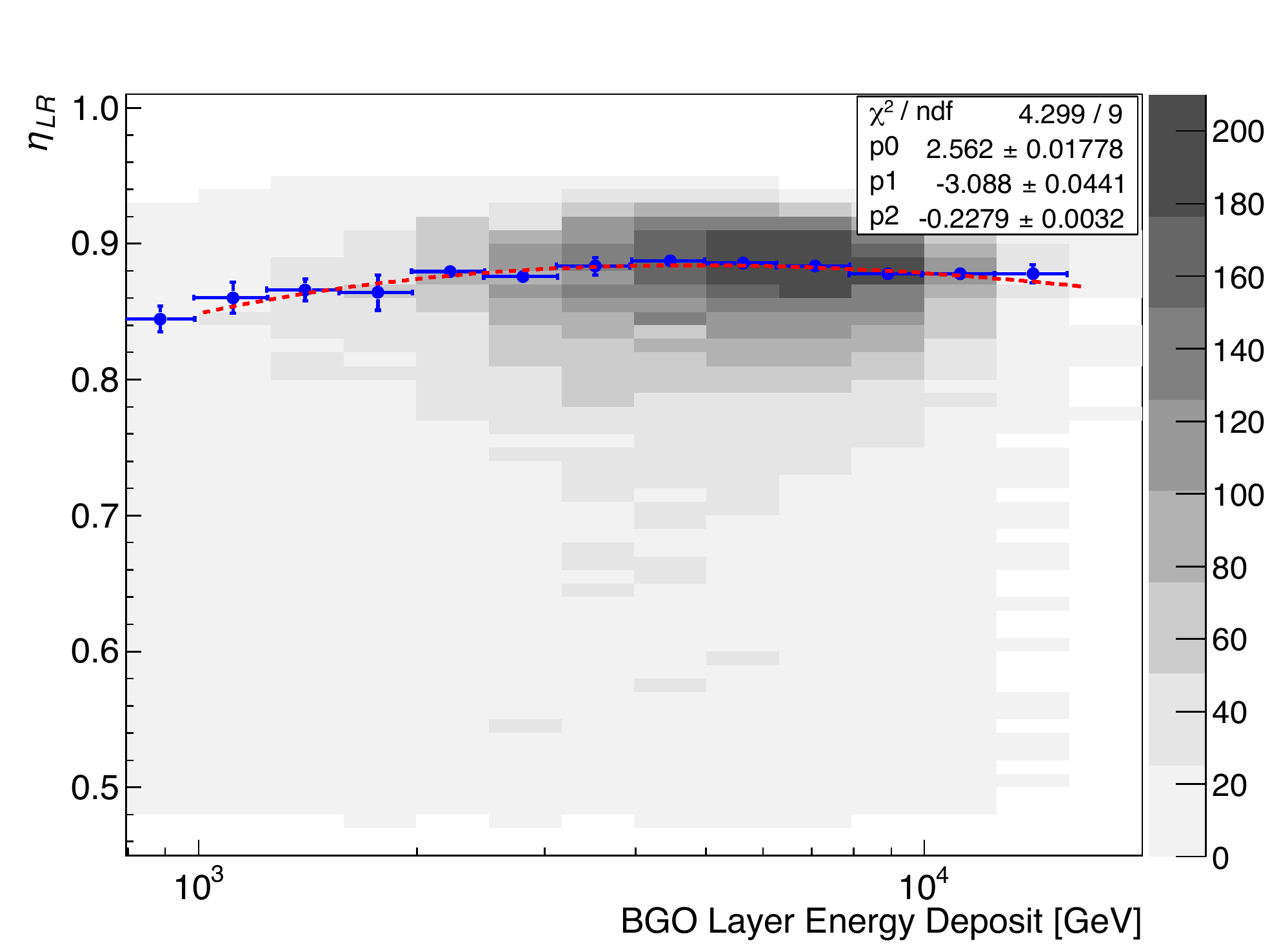}
  \includegraphics[width=.48\textwidth]{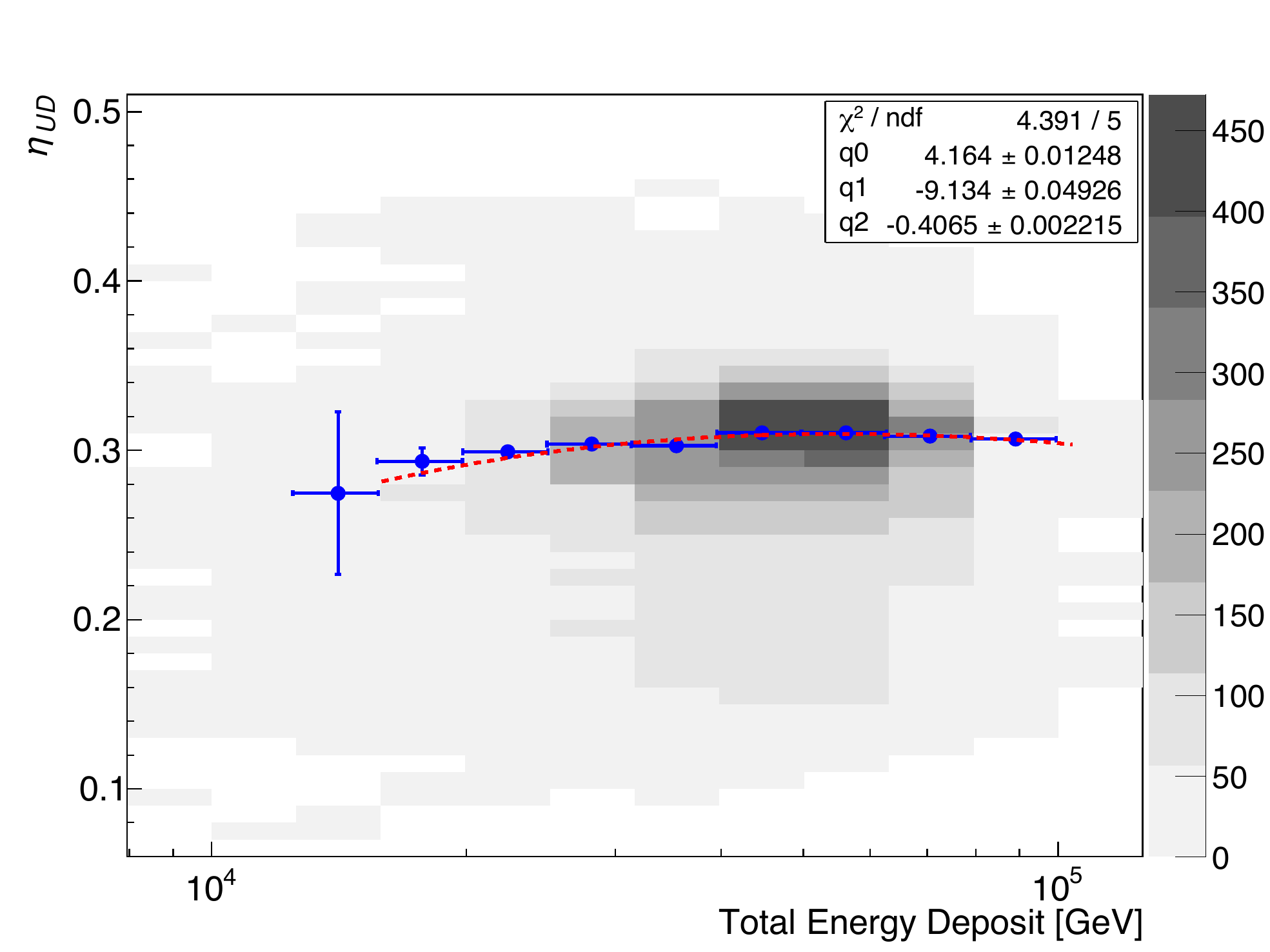}
  \caption{The profiles of $\eta_{{\rm LR}}$ of the 8th BGO layer versus the layer energy ({\tt left}) and $\eta_{{\rm UD}}$ of the 8th BGO layer versus the total deposited energy ({\tt right}) for MC helium events. The blue points and the error bars represent the fitted MPVs (most probable values) in each energy bin and their uncertainties ($\pm \sigma$) from the fit using a local gaussian function.}
  \label{fig-5}
\end{figure}

From the simulation data, we obtain the $\eta_{\rm LR}$ distribution of each layer respectively. 
In the left panel of Fig.\ref{fig-5}, the $\eta_{{\rm LR},8}$ distribution of the 8th BGO layer versus the layer energy for MC helium events is shown as an illustration. 
The profile can be fitted with an empirical function: $\eta_{{\rm LR},j} = p_{0} + p_{1}/\log(E_{{\rm layer},j}/{\rm GeV}) + p_{2} \cdot \log(E_{{\rm layer},j}/{\rm GeV})$, where $E_{{\rm layer},j}$ is the sum of energy deposits in all crystals of the $j$th layer.
The parameters $p_{0}$, $p_{1}$ and $p_{2}$ of each layer are obtained respectively. Moreover, the parameters for different nuclei, e.g. protons and helium nuclei, are obtained individually based on corresponding MC simulations.

Given the fact that the saturated crystal is the one with the maximum energy deposit in its layer, the $\eta_{{\rm LR},j}$ can be applied for saturation correction. Since the energy information of the saturated crystal is totally lost, we firstly presume an initial estimation of $E_{{\rm Sat},j} = 5.5 \cdot (E_{{\rm Left},j}+E_{{\rm Right},j})$ to calculate the $E_{{\rm layer},j}$. After that, we obtain $\eta_{{\rm LR},j}$ as the relation function of $E_{{\rm layer},j}$. With $\eta_{{\rm LR,j}}$, the energy deposit in the saturated crystal would be corrected as:
%------Equation------
\begin{equation}
E_{{\rm Sat},j} = \frac{\eta_{{\rm LR},j}}{1-\eta_{{\rm LR},j}} \cdot (E_{{\rm Left},j}+E_{{\rm Right},j}),
\label{eq-corrLR}
\end{equation}
With the updated $E_{{\rm Sat},j}$, we re-calculate $E_{{\rm layer},j}$ and $\eta_{{\rm LR,j}}$, and then apply Eq.(\ref{eq-corrLR}) once more to obtain a better estimation of $E_{{\rm Sat},j}$. For the case of more than one saturated crystals in a single shower, but existing in different layers, the correction can be performed independently for each layer. The {\tt left-right} correction is taken as the first step for the following {\tt up-down} global correction.

\subsection{Up-down correction}

After the {\tt left-right} correction, we obtain a prime energy estimation of the saturated crystal. However, to obtain a more precise energy deposit, we need to further take into account the longitudinal shower development. By considering the energy deposits of up and down layers, we construct another variable $\eta_{{\rm UD}}$, defined as
%------Equation------
\begin{equation}
\eta_{{\rm UD},j} = \frac{E_{{\rm Max},j}}{E_{{\rm Max},j}+E_{{\rm Left},j}+E_{{\rm Right},j}+E_{{\rm Up},j}+E_{{\rm Down},j}}.
\label{eq-etaUD}
\end{equation}
There are three types of definitions of $E_{{\rm Up},j}$ and $E_{{\rm Down},j}$, corresponding to classes (a), (b), and (c) in Fig.\ref{fig-4}). For case (a), $E_{{\rm Up},j}$ is defined as the sum of the maximum bar energy and the energy deposits in its left and right neighbouring bars (Sum3 for short) of the second layer, while $E_{{\rm Down},j}$ is defined as the Sum3 of the third layer. For case (b), $E_{{\rm Up},j}$ is defined as the Sum3 of the layer $j-1$, and $E_{{\rm Down},j}$ is defined as the Sum3 of the layer $j+1$. For case (c), $E_{{\rm Up},j}$ is the Sum3 of layer 12, and $E_{{\rm Down},j}$ is the Sum3 of layer 13. As an illustration, the right panel of Fig.\ref{fig-5} shows $\eta_{{\rm UD},8}$ versus the total deposited energy $E_{{\rm dep}}$ for the MC helium events. We also use the empirical form, $\eta_{{\rm UD},j} = q_{0} + q_{1}/\log(E_{{\rm dep}}/{\rm GeV}) + q_{2} \cdot \log(E_{\rm dep}/{\rm GeV})$, where $E_{\rm dep}$ is the total energy deposit in the calorimeter. As well, the parameters $q_{0}$, $q_{1}$ and $q_{2}$ are obtained individually for different layers and for different nuclei.

With the prime energy estimation of each saturated crystal after the {\tt left-right} correction, we obtain a prime estimation of the total energy deposit $E_{\rm dep}$, which is the sum of energy deposits in all crystals including the saturated one(s). By $E_{\rm dep}$, we obtain $\eta_{{\rm UD},j}$ for a further correction:
%------Equation------
\begin{equation}
E_{{\rm Sat},j} = \frac{\eta_{{\rm UD},j}}{1-\eta_{{\rm UD},j}} \times (E_{{\rm Left},j}+E_{{\rm Right},j}+E_{{\rm Up},j}+E_{{\rm Down},j}).
\label{eq-corrUD}
\end{equation}
If more than one saturated crystals exist in different layers, they would be corrected one by one globally.
The correction of Eq.(\ref{eq-corrUD}) can be performed iteratively, with updated $E_{{\rm Sat},j}$(s) and $E_{\rm dep}$. 
The results converge quickly after few iterations (three times in application).

\section{Performance}

\begin{figure}[!ht]
  \centering
  \includegraphics[width=.36\textwidth]{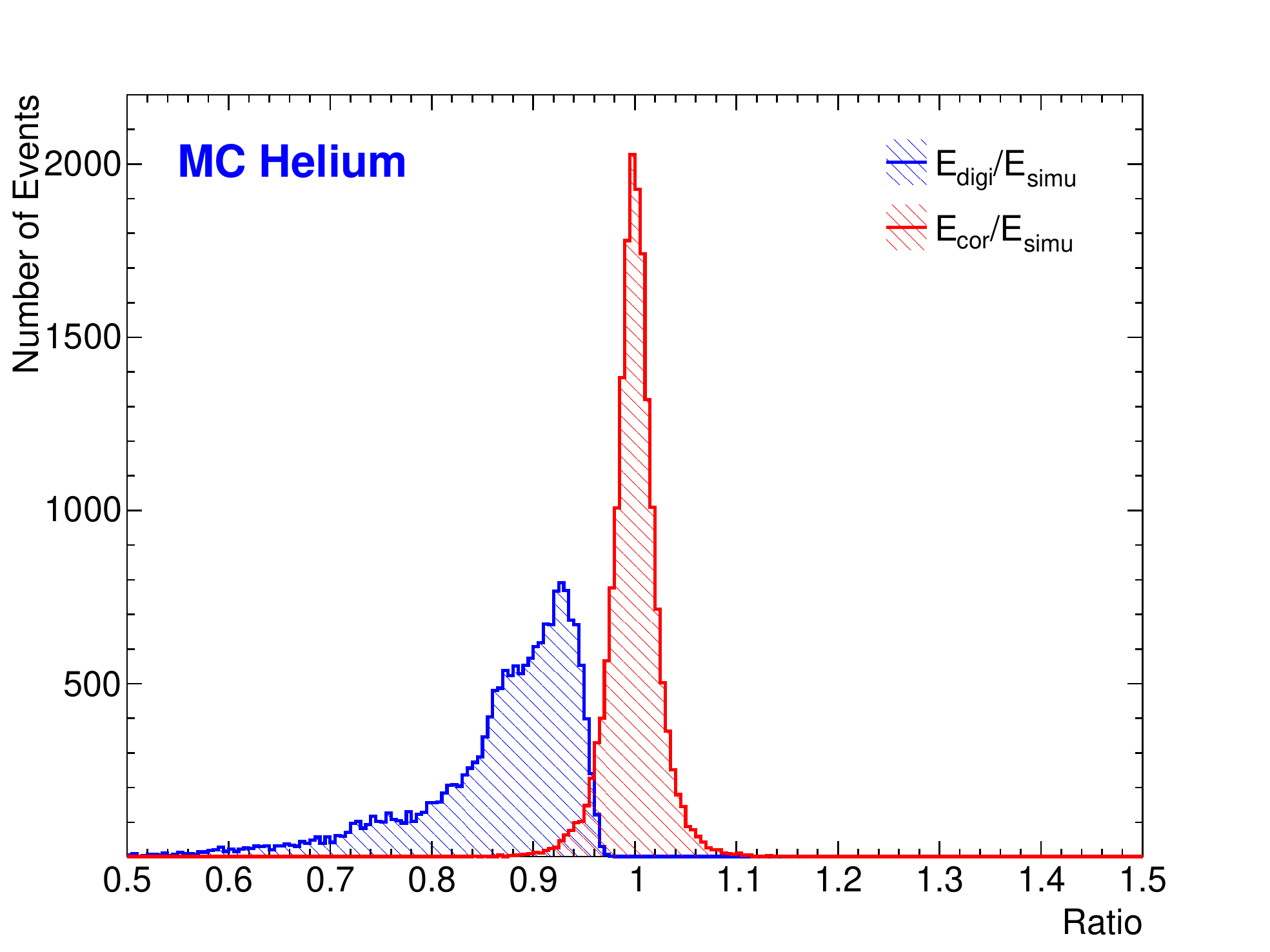}
  \put(-67,85){\color{black}{\scriptsize \tt (a)}}
  \includegraphics[width=.31\textwidth]{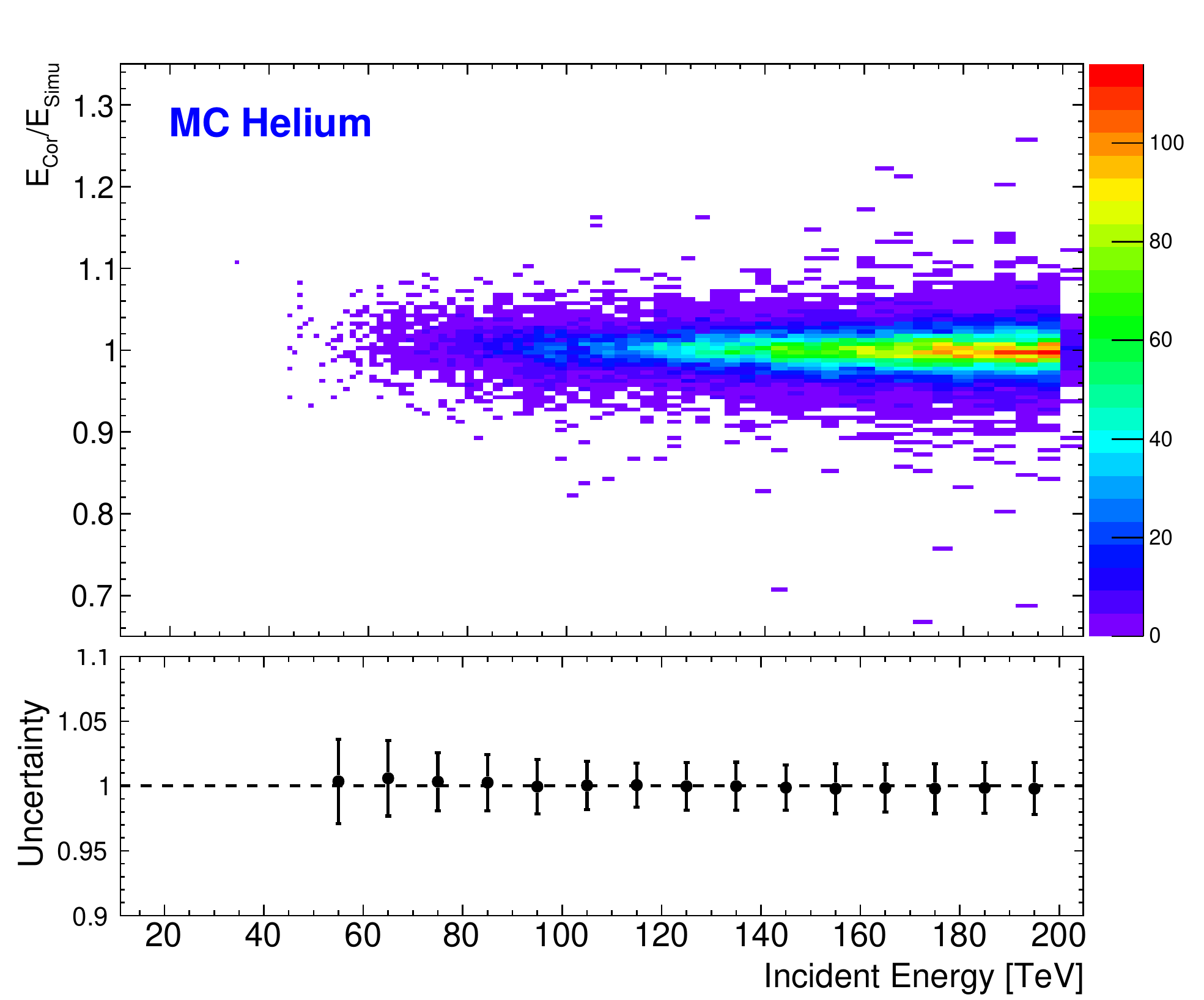}
  \put(-58,85){\color{black}{\scriptsize \tt (b)}}
  \includegraphics[width=.36\textwidth]{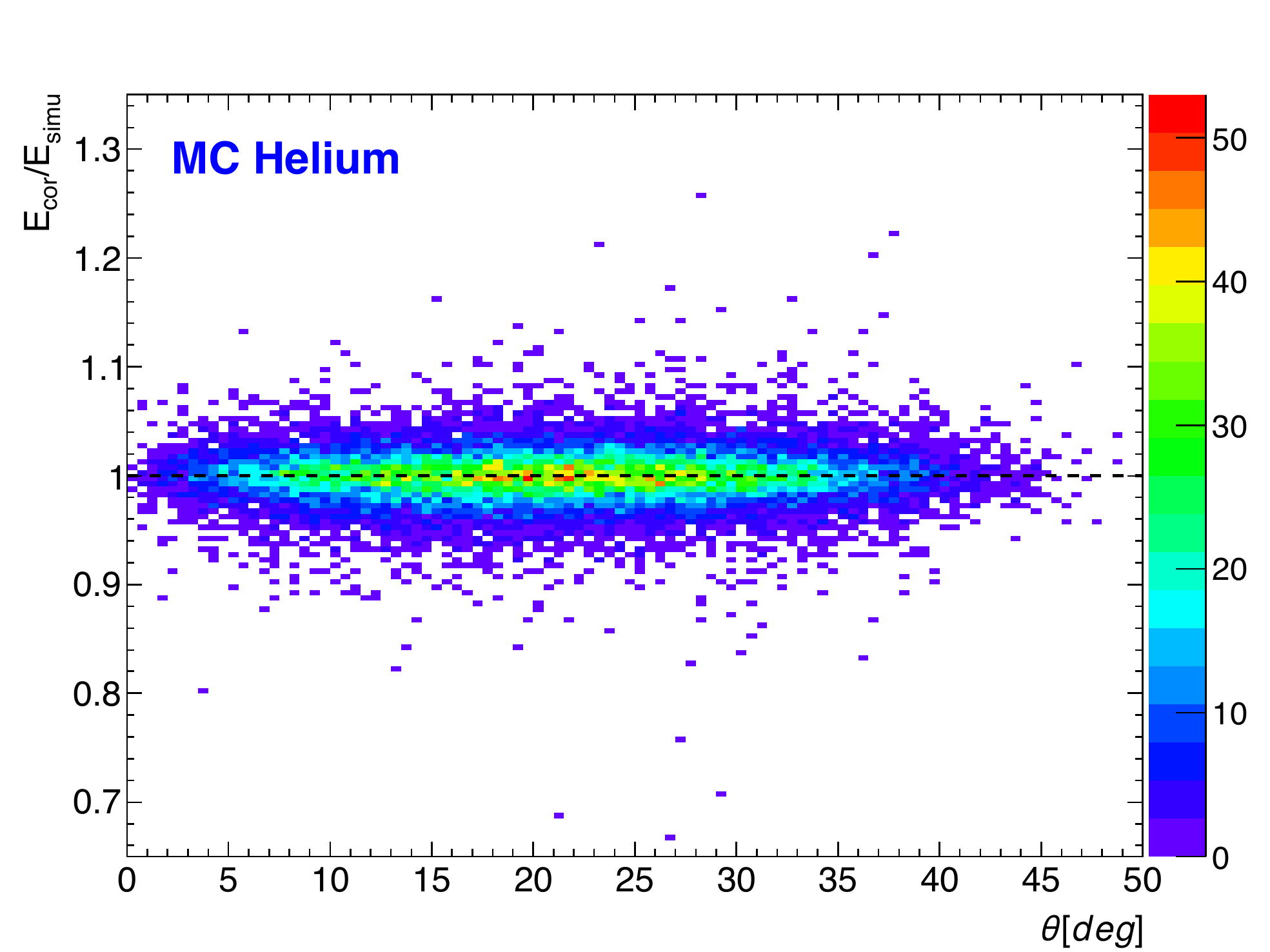}
  \put(-67,85){\color{black}{\scriptsize \tt (c)}}
  \caption{The performance of the correction method of MC saturated helium nuclei with incident energies above 10 TeV.
  {\tt (a)}: The distributions of $E_{\rm digi}/E_{\rm simu}$ (blue) and $E_{\rm cor}/E_{\rm simu}$ (red).
  {\tt (b)}: The $E_{\rm cor}/E_{\rm simu}$ ratio versus incident energy and the uncertainties ($\pm 1\sigma$) from the correction.
  {\tt (c)}: The $E_{\rm cor}/E_{\rm simu}$ ratio versus incident zenith angle $\theta$. }
  \label{fig-6}
\end{figure}

The performance of this two-step correction method is illustrated in Fig.\ref{fig-6} using high energy MC helium nuclei. The Fig.\ref{fig-6}${\tt (a)}$ shows the ratio of the corrected energy deposit ($E_{\rm cor}$) to the simulated one ($E_{\rm simu}$). The result proves that this method can well correct the energy deposits of saturated events. The $E_{\rm cor}/E_{\rm simu}$ ratios for different incident energies of MC helium data are shown in Fig.\ref{fig-6}${\tt (b)}$. We find that the performance of the correction is effective for all energies up to 200 TeV, and the uncertainty due to the correction is $\sim$2\%. In this correction method, we do not further explore the parameterizations of the correction variables, i.e. $\eta_{\rm LR}$ and $\eta_{\rm LR}$,  by considering the dependence on the incident trajectory. For one reason, the wide distributions of the correction variables are primarily due to the randomness of the hadronic shower development, rather than the incident trajectory. For another, the correction variables only have an effective dependence on the hit position for on-aixs events with a small incident zenith angle, however, the accepted particles of DAMPE are mostly oblique-incident with a zenith angle varying from 0 to 50 degree. As shown in Fig.\ref{fig-6}${\tt (c)}$, the correction is actually independent with the incident zenith angle.

\begin{figure}[!ht]
  \centering  
  \includegraphics[width=.48\textwidth]{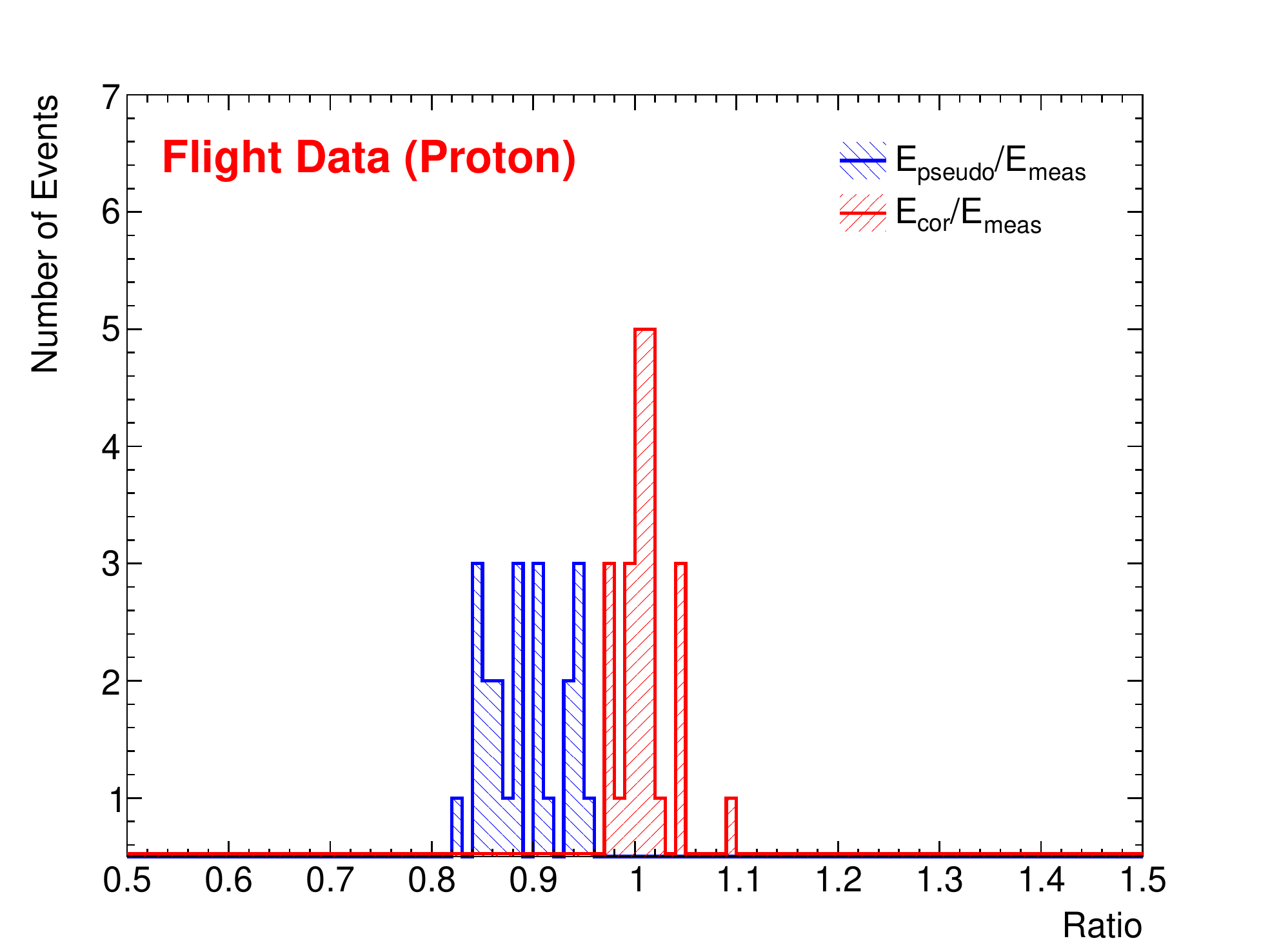}
  \includegraphics[width=.48\textwidth]{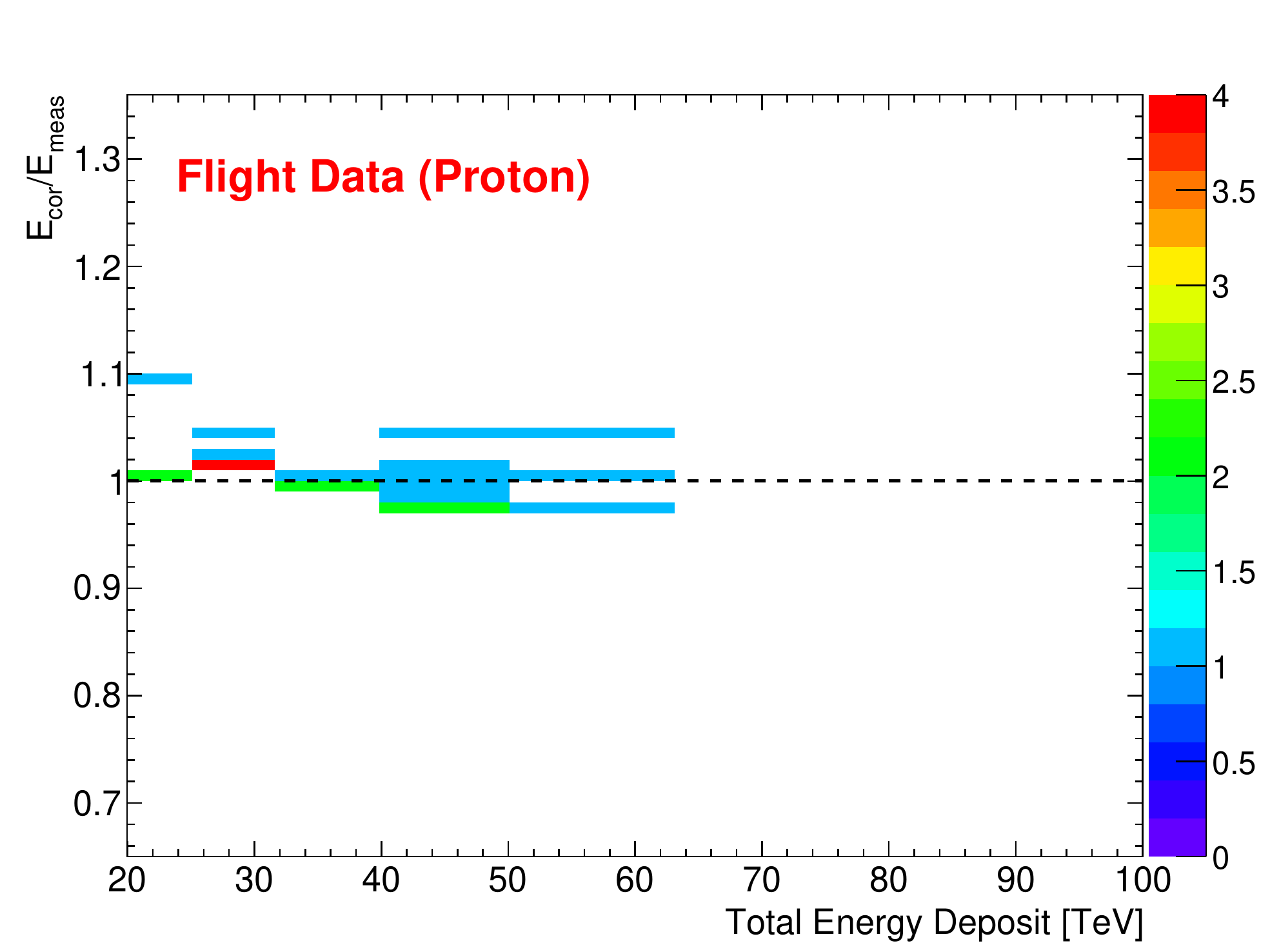}
  \caption{{\tt Left}: The distributions of $E_{\rm pseudo}/E_{\rm meas}$ (blue) and $E_{\rm cor}/E_{\rm meas}$ (red) for pseudo saturated proton candidates with total energy deposits above 20 TeV.
  {\tt Right}: The $E_{\rm cor}/E_{\rm meas}$ ratio versus total energy deposit for pseudo saturated proton candidates in the flight data.}
  \label{fig-7}
\end{figure} 

\begin{figure}[!ht]
  \centering  
  \includegraphics[width=.48\textwidth]{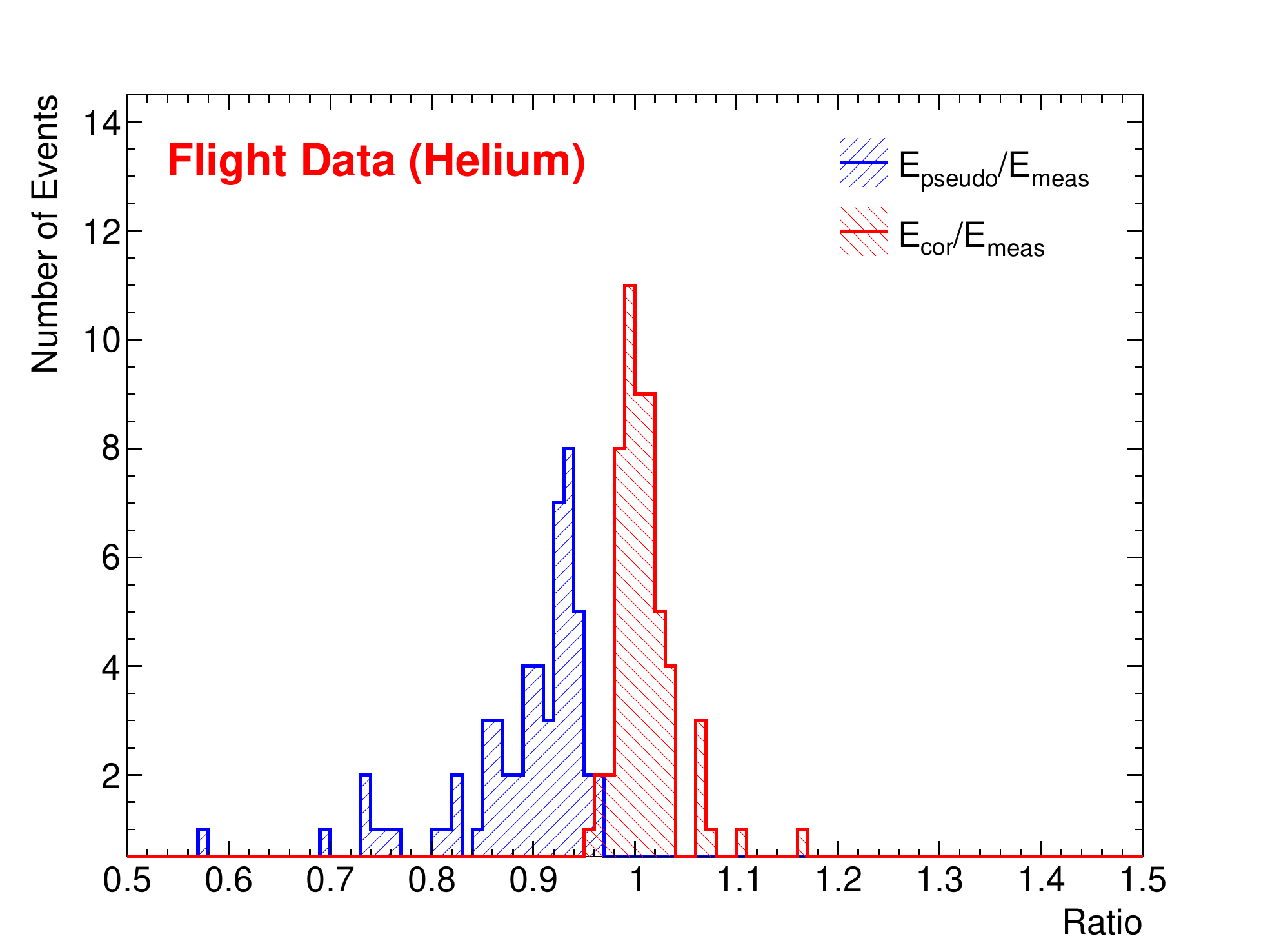}
  \includegraphics[width=.48\textwidth]{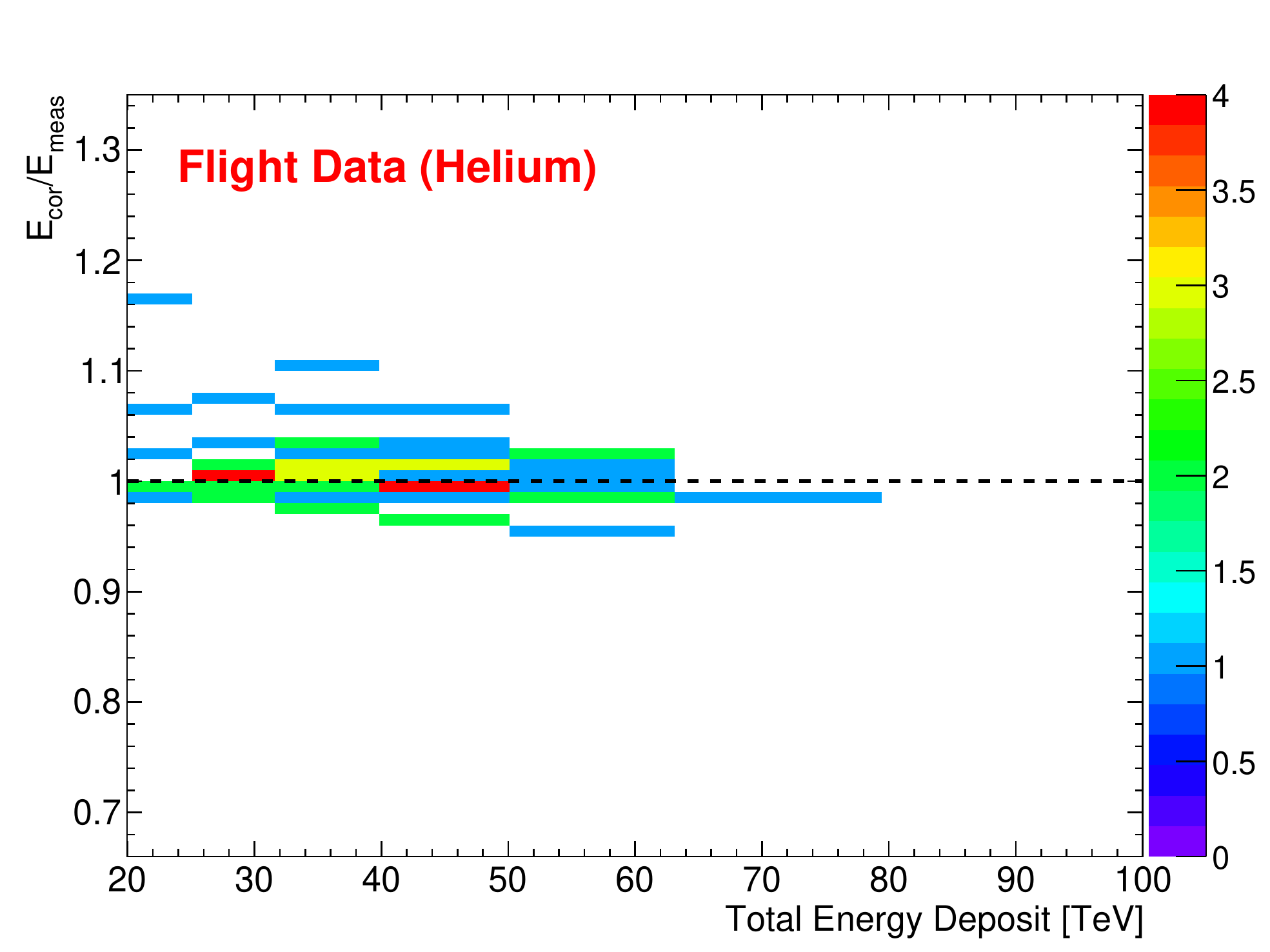}
  \caption{{\tt Left}: The distributions of $E_{\rm pseudo}/E_{\rm meas}$ (blue) and $E_{\rm cor}/E_{\rm meas}$ (red) for pseudo saturated helium candidates with total energy deposits above 20 TeV.
  {\tt Right}: The $E_{\rm cor}/E_{\rm meas}$ ratio versus total energy deposit for pseudo saturated helium candidates in the flight data.}
  \label{fig-8}
\end{figure} 

To validate the correction method with the flight data, we select high energy proton and helium candidates which are not saturated but close to the upper limit (see Fig.\ref{fig-1}). 
We require that the events should have at least one BGO crystal with energy deposits higher than $0.8 \times E_{thr}$, where $E_{thr}$ represents the measurement threshold of the corresponding crystal. 
Then we artificially remove the energy deposit(s) of such BGO crystal(s) to produce pseudo saturated events. 
The performances of the correction for the pseudo saturated proton candidates and helium candidates are shown in Fig.\ref{fig-7} and Fig.\ref{fig-8}, respectively. 
Despite the limited statistics, exported results indicate that the $E_{\rm cor}/E_{\rm meas}$ ratio shows a good independence with the total energy deposit.
For most of the pseudo saturated events, the energy deposit is properly corrected with respect to the measured one. However, it shows that a few events are slightly over-corrected. This happens because the pseudo saturated events are all under the saturation threshold, but the parameters we used for the corrections are derived with real saturated MC protons and heliums separately.

\begin{figure}[!ht]
  \centering  
  \includegraphics[width=.48\textwidth]{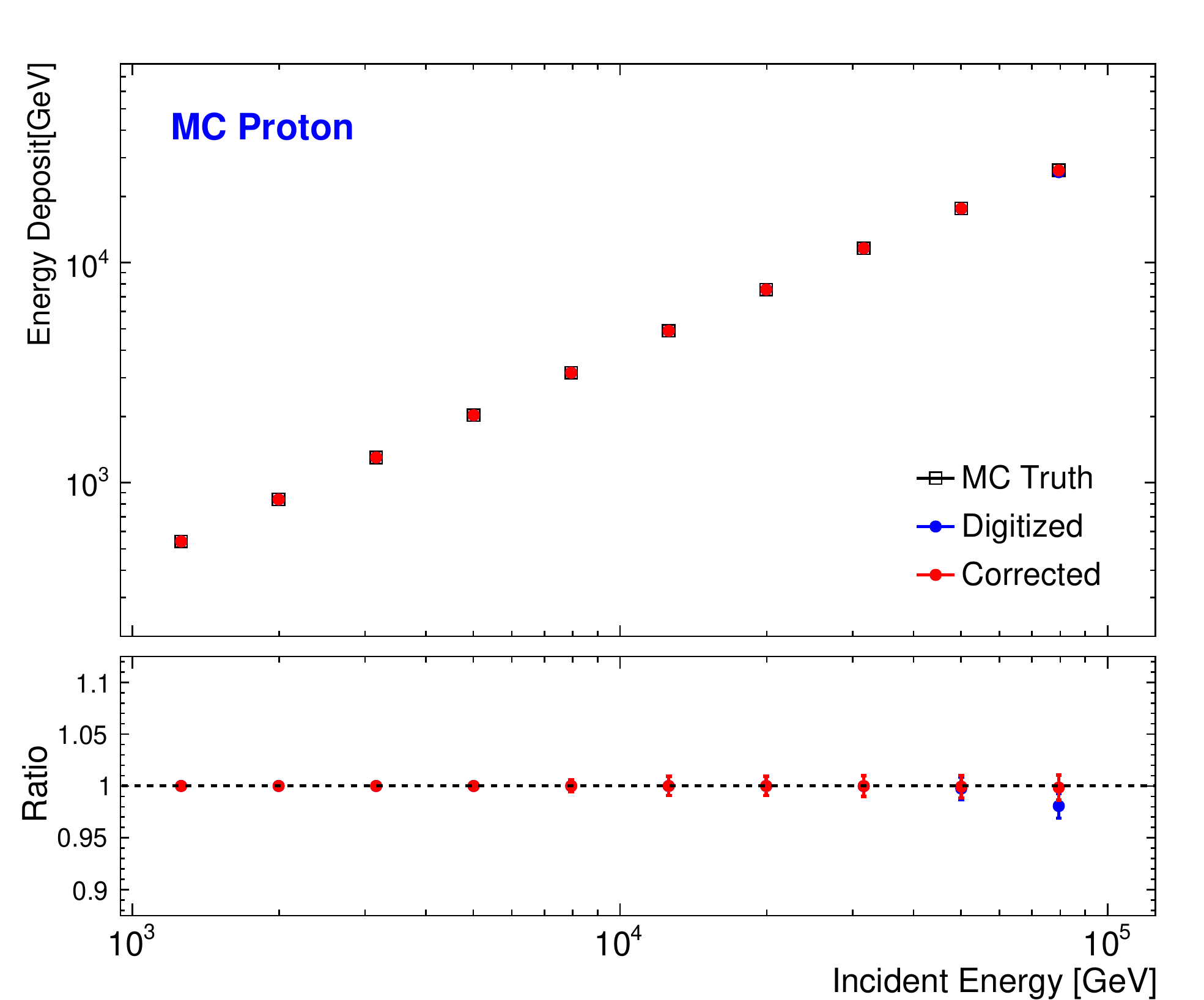}
  \includegraphics[width=.48\textwidth]{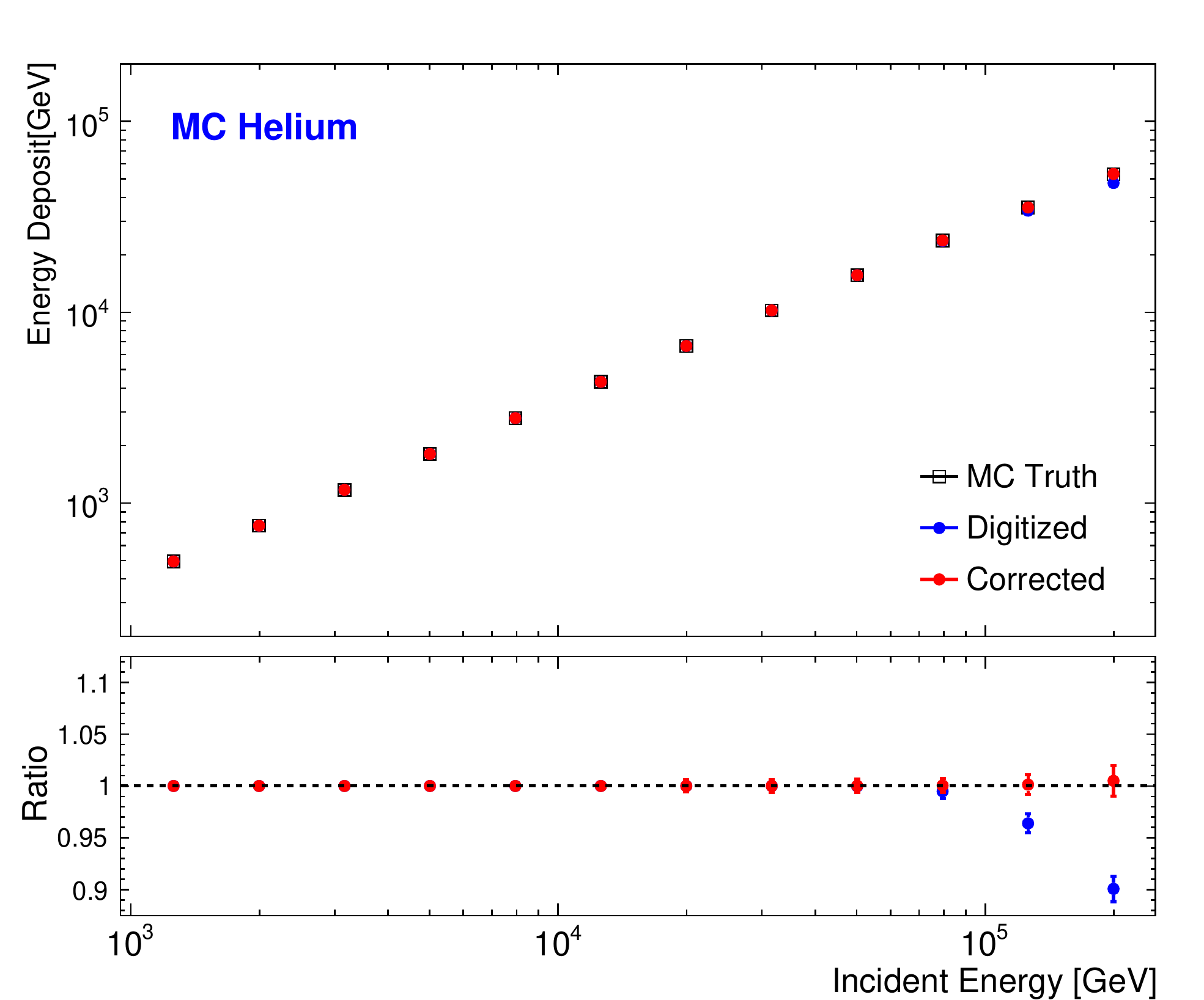}
  \caption{The digitized energies (blue dots) and corrected energies (red dots) compared with the incident energies (black squares) for MC protons ({\tt left}) and helium nuclei ({\tt right}). The bottom panels show the ratios of $E_{\rm digi}/E_{\rm simu}$ (blue dots) and $E_{\rm cor}/E_{\rm simu}$ (red dots).}
  \label{fig-9}
\end{figure}

Finally in Fig.\ref{fig-9} we show the comparisons among digitized (with saturation), corrected, and simulated energies for MC protons and helium nuclei. As can be seen in this plot, the saturation effect becomes more and more important with the increase of incident energy above 50 TeV. The correction is thus necessary for the calculation of the energy response matrix which is relevant to the spectral measurements of CR nuclei. For the proton spectrum analysis up to 100TeV in Ref.~\cite{DmpProton}, the correction has been applied for rare saturated proton candidates in the flight data. %, yet the influence for the final spectrum is negligible. 
%In the spectrum analysis, one of the most important issues is the response matrix construction \cite{DmpProton}. Fig.\ref{fig-7} shows the mean value of total energy deposit along with the incident energy for MC helium nuclei. With the saturation effect included in the digitization, the digitized energy deposit deviates from the MC truth and the deviation becomes larger with the increase of incident energy above decades of TeVs. After applying the correction, the energy deposit is consistent with the MC truth.

\section{Conclusions}
In order to extend the energy measurements of the DAMPE for hadronic CRs to sub-PeV energy ranges, the BGO readout saturation effect has been studied based on detailed MC simulation data. 
%The BGO readout saturation effect will underestimate the total energy deposit in the calorimeter, thereby biasing the real energy response for high energy hadronic cosmic-rays. 
Through combining the energy information of neighbouring BGO crystals and the longitudinal shower development, we proposed a two-step correction method to reconstruct the energy deposit of saturated crystals. The first step is to use the {\tt left} and {\tt right} energy deposits of the saturated crystal to get a prime estimation of the saturated crystals. Then the longitudinal shower development is further taken into account to improve the correction. The correction parameters are obtained for different nucleonic species. The performance of the correction method is illustrated using MC helium nuclei and also helium candidates in flight data, which show that the energy deposits of saturated crystals can be well reconstructed. The correction is expected to be very helpful in the measurements of the CR spectra at very high energies.

One caveat of the correction method is that it applies only for the case with no adjacent saturated crystals within the same layer. The events with two or more adjacent crystals of the same layer get saturated are very rare, but existing in the flight data. The correction for such events would be more complicated and uncertain. We leave such a study in future works.

\section{\it Acknowledgments.}
%The authors would like to thanks the colleagues of DAMPE collaboration for useful discussions. 
This work is supported by the National Key Research and Development Program of China (Grant No. 2016YFA0400200), and the National Natural Science Foundation of China (Grant Nos. 11722328, 11773085, U1738205, U1738207, 11851305), and the 100 Talents Program of Chinese Academy of Sciences. 
%In Europe, the analysis are supported by the Swiss National Science Foundation (SNSF), Switzerland; the National Institute for Nuclear Physics (INFN), Italy.

\bibliography{mybib}
\bibliographystyle{elsarticle-num}

\end{document}